\documentclass[pdflatex,sn-mathphys-num]{sn-jnl}% Math and Physical Sciences Numbered Reference Style
\usepackage{graphicx}%
\usepackage{multirow}%
\usepackage{amsmath,amssymb,amsfonts}%
\usepackage{amsthm}%
\usepackage{mathrsfs}%
\usepackage[title]{appendix}%
\usepackage{xcolor}%
\usepackage{textcomp}%
\usepackage{manyfoot}%
\usepackage{booktabs}%
\usepackage{algorithm}%
\usepackage{algorithmicx}%
\usepackage{algpseudocode}%
\usepackage{listings}%
\usepackage[per-mode = symbol]{siunitx}
\usepackage{bbm} %---bra, ket definition %
\DeclareSIUnit{\amagat}{amg}
\DeclareSIUnit{\sample}{Sa}
\usepackage{xcolor}
\usepackage{color}
\definecolor{mygreen}{rgb}{0,0.5,0}
\definecolor{mygrey}{rgb}{0.5,0.5,0.5}
\definecolor{myred}{rgb}{0.75,0,0}
\definecolor{myblue}{rgb}{0,0,0.75}
\definecolor{mymagenta}{cmyk}{0,1,0,0.12}
\definecolor{mycyan}{cmyk}{1,0,0,0.12}
\definecolor{myorange}{rgb}{1.,0.5,0}
\definecolor{myviolet}{rgb}{0.6,0.15,0.6}
\definecolor{mybrown}{cmyk}{0,0.50,1,0.41}

\usepackage{ulem}

\newcommand{\BDC}{B_\mathrm{DC}}

\newcommand{\subMTB}{_\mathrm{MTB}}

\newcommand{\Bpol}{B_\mathrm{pol}}
\newcommand{\Mfull}{M_\mathrm{full}}

\newcommand{\myhat}[1]{\hat{\mathbf{#1}}}

\newcommand{\yhat}{\myhat{y}}
\newcommand{\zhat}{\myhat{z}}
\newcommand{\rhat}{\myhat{r}}

\usepackage{mathtools}
\newcommand{\Dtwo}{\ensuremath{\mathrm{D}_2}{}}
\newcommand{\Done}{\ensuremath{\mathrm{D}_1}{}}
\renewcommand{\Dtwo}{\relax\ifmmode\mathrm{D}_2\else D\textsubscript{2}{}\fi}
\renewcommand{\Done}{\relax\ifmmode\mathrm{D}_1\else D\textsubscript{1}{}\fi}

\newcommand{\tauAllan}{\tau}
\newcommand{\tauBact}{\tau_B}

\newcommand{\PDone}{\relax\ifmmode\mathrm{PD}_1\else PD\textsubscript{1}{}\fi}
\newcommand{\PDtwo}{\relax\ifmmode\mathrm{PD}_2\else PD\textsubscript{2}{}\fi}

\DeclareSIUnit{\dBm}{dBm}
\DeclareSIUnit{\torr}{Torr}

\DeclareMathOperator{\var}{var}
\newcommand{\ICFO}{ICFO - Institut de Ci\`encies Fot\`oniques, The Barcelona Institute of Science and Technology, 08860 Castelldefels (Barcelona), Spain}
\newcommand{\ICREA}{ICREA - Instituci\'{o} Catalana de Recerca i Estudis Avan{\c{c}}ats, 08010 Barcelona, Spain}
\newcommand{\BARI}{Dipartimento Interateneo di Fisica, Universit\'{a}
degli Studi di Bari Aldo Moro, 70126 Bari, Italy}

\theoremstyle{thmstyleone}%
\theoremstyle{thmstyletwo}%

\theoremstyle{thmstylethree}%

\begin{document}

\title[Article Title]{Magnetotactic bacterial populations studied with a Pound-Drever-Hall atomic magnetometer 
%\\ \otext{Magnetotactic bacterial populations studied with a Pound-Drever-Hall atomic magnetometer}
}

\author*[1]{\fnm{Mar\'{i}a Hern\'{a}ndez Ruiz} }\email{maria.hernandez@icfo.eu}

\author[1]{\fnm{Christopher Kiehl} }

\author[1,2]{\fnm{Vito Giovanni Lucivero}}
\author[1,3]{\fnm{Morgan W. Mitchell} }

\affil*[1]{\ICFO}

\affil[2]{\BARI}

\affil[3]{\ICREA}


\abstract{We demonstrate an optically pumped magnetometer that monitors spin polarization using Pound Drever Hall (PDH) technique. The instrument exhibits a noise floor of \SI{22.2}{\pico\tesla\per\sqrt\hertz} limited by optical photon shot noise, short-term instability of $(\SI{30.8}{\pico\tesla\per\sqrt\hertz} )/\sqrt{\tauAllan}$ for averaging times  $\tauAllan \le \SI{0.2}{\second}$, instability below \SI{70}{\pico\tesla} for $\SI{0.2}{\second} \le \tauAllan \le \SI{20}{\second}$ and a  minimum instability of \SI{47}{\pico\tesla} at $\tauAllan = \SI{6}{\second}$. We apply the OPM to investigate the ability of magnetotactic bacteria (Magnetospirillum gryphiswaldense, MSR-1) to orient in externally applied magnetic fields. Observing an opaque, concentrated suspension, we detect deviations from exponential relaxation dynamics on second time-scales, which give information about the dispersion of bacterial magnetic moment and rotational damping coefficient. These parameters are observed to evolve as the population further concentrates due to evaporation and settling. To our knowledge, this is the first time such magnetic inhomogeneities and long-term relaxation deviations have been directly observed.
This study showcases both the sensitivity and stability of our OPM and its potential for probing biophysical processes. }




\maketitle

\section{Introduction} \label{sec:Introduction} 

Optically-pumped magnetometers (OPMs) \cite{OpticalMagnetrometry, FabricantNJP2023} offer high sensitivity and absolute accuracy for applications ranging from magnetoencephalography \cite{BotoNI2017, BotoN2018, TierneyNeuroimage2019} to space exploration \cite{ArridgeNP2016, DoughertyS2006} to searches for physics beyond the standard model \cite{SafronovaRMP2018}. For any given atomic species, the fundamental limits of OPM sensitivity scale inversely with the volume of vapor probed \cite{JimenezBook2017, MitchellRMP2020}. Consequently, most high-sensitivity OPMs use centimeter- or millimeter-dimension vapor cells, with corresponding limits on spatial resolution. There is, nonetheless, interest in pushing OPM imaging to sub-mm spatial resolutions \cite{YangPRAp2017, KnappePatent2024,rasser2025towards}, i.e., into the domain of magnetic microscopy \cite{CHEMLA19993323,NVBACT}. This could open new applications in materials science \cite{FeketeNL2024, YangNP2020}, in fundamental physics, and in biomagnetism of tissues and small organisms. One challenge in developing OPMs with very small working sensing volumes is the consequent reduction in optical depth, which implies a less efficient readout of the atomic signal by, e.g., Faraday rotation probing \cite{HernandezPRAppl2024}. An appealing solution to this challenge is to resonate the probe light within an optical cavity \cite{CrepazSR2015, MazzinghiOE2021}, to enhance the optical extraction of signal from the atomic medium. Other techniques such as multipass probing~\cite{li2011optical} can increase optical depth but face increasingly stringent fabrication and alignment requirements as the cell dimensions decrease. In addition, the number of effective passes in a multipass geometry is fundamentally constrained by the beam’s Rayleigh range for a given cell size, and achieving uniform illumination across the vapor cell volume remains a challenge~\cite{Kee2025SpinCorrelations}.

Here we describe the application of an OPM with cavity-based optical readout to measurement of magnetic relaxation signals from the magnetotactic bacteria (MTB) \cite{RevModPhys.96.021001} \textit{Magnetospirillum gryphiswaldense}, strain MSR-1 \cite{BacteriaUPV}.  These microorganisms biomineralize chains of iron oxide nanocrystals to form magnetosomes, which impart a permanent magnetic dipole moment to the bacterium  and enable alignment with external magnetic fields~\cite{barber2016invagination}. MTBs are of particular interest not only for their potential biomedical applications, such as targeted tumor therapy via magnetic hyperthermia~\cite{BacteriaUPV}, but also as model systems for exploring active matter~\cite{petroff2025magnetotactic,birjukovs2025magnetic} and magnetohydrodynamics~\cite{thery2020self}. Magnetometer techniques complement optical MTB measurement ~\cite{rosenblatt1982birefringence, zhao2007simple, lefevre2009characterization, welleweerd2022open}, as they can study MTB in bulk and in opaque conditions (due, e.g., to a strongly-scattering environment, opaque container, or simply due to a high concentration of MTB themselves), and offer direct information of the magnetic response, independent of bacterial morphology \cite{lefevre2009characterization}.  Enhancing the spatial resolution and sensitivity of magnetometers could enable meaningful bulk measurements at lower MTB concentrations and even extend to single-cell resolution in a magnetic microscopy configuration, as has previously been demonstrated with superconducting quantum interference devices (SQUIDs)~\cite{CHEMLA19993323}. 
%Such capabilities would complement optical microscopy approaches to single-MTB studies. While SQUIDs show great promise for biomagnetic measurements, OPMs offer distinct practical advantages by providing high sensitivity without the need for cryogenic cooling.

The OPM we use here implements a free-induction decay (FID) scalar magnetic measurement, with resonantly-enhanced readout using a planar external optical cavity and Pound-Drever-Hall (PDH) detection \cite{HernandezPRAppl2024}.   Even with concentrated MTB samples ($\sim \SI{1e11}{cells\per\milli\liter}$), the biomagnetic signal is of $\sim \SI{}{\nano\tesla}$ magnitude, and relaxes to a thermal-magnetic equilibrium on $\sim \SI{10}{\second}$ time scales as the bacteria re-orient, in the presence of a $\sim \SI{50}{\micro\tesla}$ Earth-field background \cite{RevModPhys.96.021001}. As we show, these slow, weak signals can nonetheless be measured with sub-\SI{100}{\pico\tesla} precision by the PDH-OPM, due to its high sensitivity -- comparable to earlier work with a dcSQUID microscope %also with sub-\SI{100}{\pico\tesla} precision 
\cite{CHEMLA19993323} --  and high stability in Earth field conditions. The PDH-OPM is built around a \SI{}{\milli\meter}-scale microfabricated vapor cell (MEMS cell), showing compatibility with techniques necessary for OPM magnetic microscopy, and in an one-layered open shield, to allow rapid access to the biological samples. The OPM performance under these conditions shows that meaningful MTB relaxation signals can be recovered without the expense and inconvenience of strong magnetic isolation. Finally, we show the PDH-OPM can quantify small deviations from exponential relaxation that cause parts-per-million changes in the observed field strength. In MTB populations such deviations arise from dispersion of the magnetic moment and viscous damping coefficient. Using this capability we observe immobilization of a fraction of the MTB population as the solution is concentrated by evaporation. 

The article is organized as follows. Section~\ref{Sec:Experimental system} presents the experimental system. Section~\ref{sec:MagnetometerSensitivity} details the working principle and sensitivity of the magnetometer. In Section~\ref{sec:BacteriaResults}, we present measurements of the magnetic signal arising from bacterial alignment in an applied magnetic field, while Section~\ref{sec:Distributions} analyzes changes in the exponential behavior of the magnetotactic bacteria signal in the presence of sample evaporation.
\begin{figure}[h]
\centering
\includegraphics[width=0.94\linewidth]{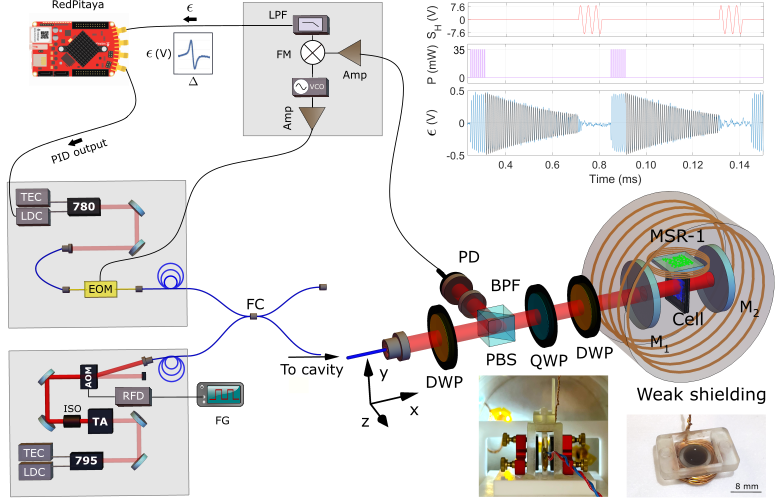}
    \caption{Experimental setup. Pump and probe beams are coupled into a fiber coupler (FC) and reach the atomic vapor within the cavity in the same collinear mode. The probe light is modulated in phase, the pump light is amplified and modulated in amplitude. Pumping scheme is shown with purple continuous line in the diagram. The reflected probe is collected by a 1.4-GHz bandwith  photodetector PD, whose output is fed into a Pound-Drever-Hall (PDH) circuit giving an output error signal $\epsilon$. When locking to the zero-crossing point,  $\epsilon$ follows atomic precession, as show in the diagram above, where blue line is raw data and black line a fitting to Equation~\ref{Eq:FID}.  $S_H$ in red is the voltage sent to the heater.  There are two coils to generate magnetic fields in two perpendicular directions, the shielding coil and the polarizing coil. In the bottom, picture of the optical cavity in the shield with one cap open. The 3D-printed structure is glued to the base of the cavity using epoxy. Positioned above this is the bacteria holder, which incorporates the polarizing coil. This holder is not permanently fixed to the structure, allowing for easy removal and portability. On the right, a photograph displays the opened, bacteria holder containing the solution.    LDC - laser diode current controller, TEC - laser diode temperature controller, EOM - electro-optic modulator, TA - tapered amplifier, ISO - optical isolator, AOM - acousto-optic modulator, RFD - radio frequency driver, FG - function generator, VCO - voltage-controlled oscillator, FM - frequency mixer, LPF - low-pass filter, DAQ - digital oscilloscope. Sketch of the cavity with detection in reflection. DWP - dual wavelength wave-plate, PBS - polarizing beam splitter, QWP - quarter wave-plate, BPF - laser line bandpass filter, PD- photo-detector, M\textsubscript{1} / M\textsubscript{2} - first/second planar cavity mirror. }  \label{fig:setup}
\end{figure}

\section{Experimental system}
\label{Sec:Experimental system}

The experimental system is illustrated schematically in Figure~\ref{fig:setup} and described briefly in this section. Full  details are given in Hernandez \textit{et al.} \cite{HernandezPRAppl2024}. 
\textsuperscript{87}Rb vapor with \SI{1.3}{\amagat} of N\textsubscript{2} buffer gas  is housed in a MEMS cell, inside a Fabry-Perot planar optical resonator  formed by two parallel planar mirrors spaced along the $x$ direction (the cavity axis). The interior length of the cell is \SI{1.5}{\milli\meter}. The vapor's spin polarization along this axis is continuously monitored through the shift in the cavity resonance which we detect using PDH detection: a  $\SI{5}{\milli\watt}$ continuous-wave (CW) laser blue-detuned $\SI{66}{\giga\hertz}$  from  the \SI{780}{\nano\meter} D\textsubscript{2} transition is phase modulated, reflected from the optical cavity and detected on a photo-detector (PD). The  PD output signal is demodulated to obtain a PDH error voltage $\epsilon$, which, near cavity resonance, is proportional to the cavity line shift induced by the $x$-component of the spin polarization. This PDH signal is continuously monitored to obtain records as those shown in Figure~\ref{fig:setup}.

Spin polarization is generated by Bell-Bloom excitation: a  CW diode laser and tapered amplifier, followed by an  acousto-optic modulator, are used to produce trains of circularly-polarized optical pulses, each resonant to the \SI{795}{\nano\meter} D\textsubscript{1} transition. The train consists of square pulses with a  $\SI{30}{\percent}$ duty cycle and a pulse-repetition period equal to the nominal Larmor period. As seen in Figure~\ref{fig:setup}, sufficient pulses are applied to saturate the atomic vapor polarization. 

 The FID sequence is repeated every $t_0 = \SI{600}{\micro\second}$ and contains three different phases: a pumping phase, a probing phase and heating dead-time phase.   First, atomic spins are polarized as just described for $\sim\SI{68}{\micro\second}$. Then, they are allowed to freely precess around the magnetic field and the PDH signal $\epsilon$, which indicates the axial spin polariation, recorded as a function of time, see Figure~\ref{fig:setup}. A brief period of alternating-current (ac) Joule heating is then applied for \SI{100}{\micro\second}: A resistive heater surrounding the MEMS cell is driven by three cycles of ac current at frequency \SI{30}{\kilo\hertz}. The voltage applied to the heater $S_H$ is shown in Figure~\ref{fig:setup}. Due to this heating, the MEMS cell equilibrates to a mean temperature of $\approx \SI{127}{\celsius}$, and thus a number density of $n_A\approx 2.9\times10^{13}$ atoms/cm$^3$ , as estimated from the cavity transmission spectrum \cite{HernandezPRAppl2024}. 

A sample holder,  composed of two distinct 3D-printed components, was fabricated using FormLabs Clear Resin v4.1. The first component, which serves as the base, has a height of \SI{37.6}{\milli\meter} and is permanently fixed in place using epoxy. It provides stable support for the removable upper part of the holder. The second component, which is inserted into this base, contains a cylindrical compartment, in which we place the MTB solution. This compartment has a diameter of \SI{8}{\milli\meter} and a height of \SI{4.5}{\milli\meter}. The bottom wall has a thickness of \SI{0.5}{\milli\meter}, allowing the center of the sample to be positioned at a distance of \SI{17}{\milli\meter} from the atomic cell center.  Surrounding the bacterial sample, there is a polarizing coil that generates a magnetic field along the $y$-axis of \SI{1}{\milli\tesla} to initially polarize the bacteria along this direction. A solenoid coil wound at the inner surface of the shield produces a uniform field of up to \SI{150}{\micro\tesla}  along $z$. This field 1) produces an Earth-field-scale background in which the MTB  become polarized on $\sim \SI{1}{\second}$ time-scales and 2) provides a finite offset field that determines the  Larmor frequency for FID and stroboscopic optical pumping. The MTB contribute a small field that reinforces or diminishes this offset field.

The bacterial strain used is \textit{Magnetospirillum gryphiswaldense} (MSR-1). These are immobilized in distilled water by the addition of glutaraldehyde to a final concentration of \SI{2}{\percent}. The initial bacterial concentration in the solution is \SI{1e11}{cells\per\milli\liter}  and we use \SI{120}{\micro\liter} of the solution. Due to thermal convection from the cell's oven, the bacteria solution reaches a temperature of \SI{32}{\celsius}.

\begin{figure}[t]
    \centering
    \includegraphics[height=3.6 cm]{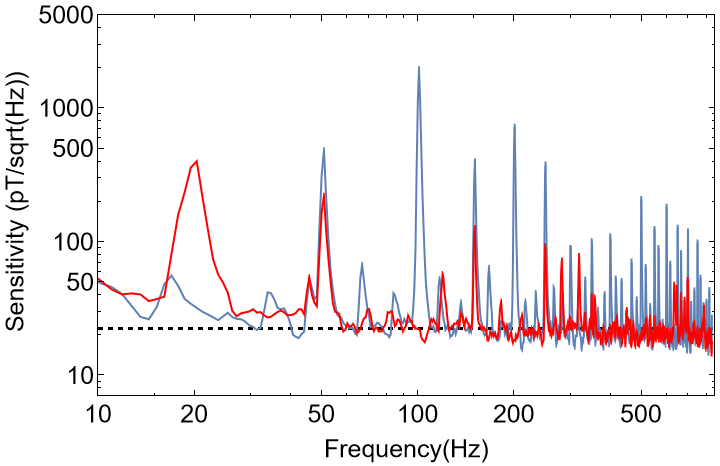} 
    \hspace{0.2 cm}
    \includegraphics[height=3.6 cm]{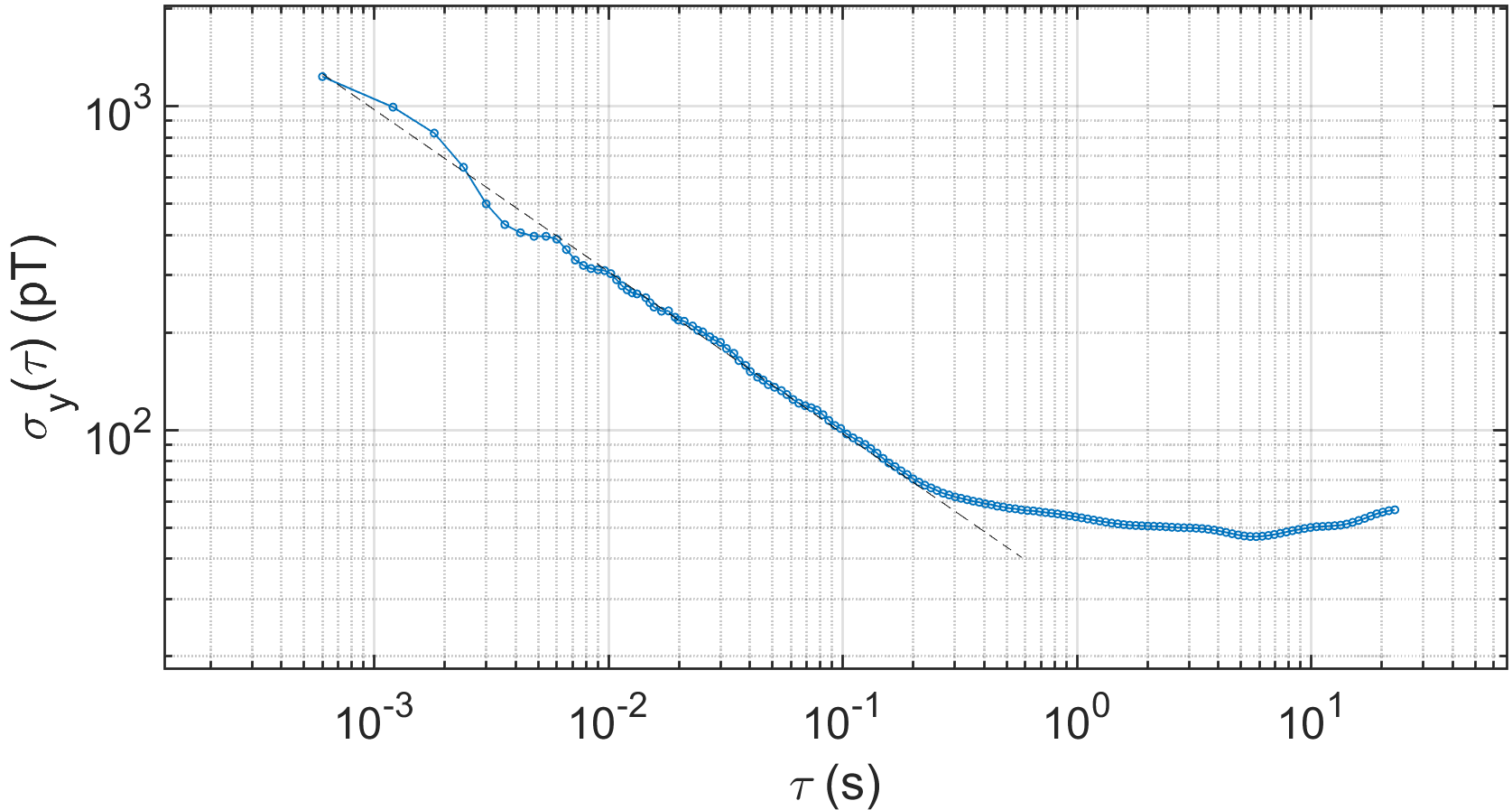}
    \caption{Magnetometer sensitivity and stability. Left: measured OPM equivalent magnetic noise (Amplitude spectral density ) for applied DC fields of \SI{17}{\micro\tesla} (red) and \SI{147}{\micro\tesla} (blue) along the $z$ direction.  Dashed line indicates a noise floor of \SI{22.2}{\pico\tesla\per\sqrt\hertz}. Spectra averaged 10 times. 
    Right: overlapping Allan deviation (OAD). With an applied field of \SI{17}{\micro\tesla}, a single acquisition of \SI{22.6}{s} duration, filtered with fourth-order, \SI{2}{Hz}-wide IIR bandstop filters at \SI{50}{Hz}, \SI{100}{Hz}, \SI{150}{Hz} and \SI{200}{Hz}. Dashed line shows the OAD of uncorrelated white noise: $\sigma = (\SI{30.8}{\pico\tesla\per\sqrt\hertz} ) \tauAllan^{-1/2}$.}
    \label{fig:MagneticSensAndStab}
\end{figure}

\section{ Magnetometer sensitivity and stability}
\label{sec:MagnetometerSensitivity}

A short sequence of representative FID traces is shown in Figure~\ref{fig:setup}, together with fits with the FID function 
\begin{equation}
\label{Eq:FID}
\epsilon(t)=P_0 e^{-t/T_2} \sin(\omega_\mathrm{L} t+\phi),
\end{equation}
where $P_0$ is and amplitude proportional to the initial polarization, $\omega_\mathrm{L}$ is the linear Larmor precession frequency, $\phi$ is the phase and $T_2$ is the transverse relaxation time. From such fits we obtain an estimate of the field strength  $B = \omega_\mathrm{L}/ \gamma$  , where  $\gamma = 2\pi \times \SI{7}{\hertz\per\nano\tesla}$ is the \textsuperscript{87}Rb gyromagnetic ratio, every \SI{600}{\micro\second}. Figure~\ref{fig:MagneticSensAndStab} shows the measured magnetic sensitivity (amplitude spectral density) for two nominally-constant field strengths ~of $\SI{17}{\micro\tesla}$ and $\SI{147}{\micro\tesla}$, each obtained by Fourier transform the magnetic field data and averaging ten sections of $\SI{1.17}{\second}$ duration each.

To understand the OPM stability, we collect $M$ samples of the field $B(t)$ with sampling period $t_0$, i.e., with $t=j t_0$, $j \in \{1, \ldots , M\}$, and calculate the overlapping Allan variance as
\begin{equation}
    \sigma^2_y(\tauAllan)=\frac{1}{2(M-2n+1)} \sum_{i=1}^{M-2n+1}{(y_{i+n}^{(n)}-y_i^{(n)})^2},\label{eq:allan2}  
\end{equation}
where $\tauAllan = n t_0$, $n\in \mathbbm{N}$ is the averaging time, and
\begin{equation}
    y_{i}^{(n)} \equiv  \frac{1}{n} \sum_{j=i}^{i+n-1} B(j t_0)
\end{equation}
is $B(t)$ averaged over the $i$th time window \cite{ZhangM2008}. As shown in Figure~\ref{fig:MagneticSensAndStab}, the overlapping Allan deviation $\sigma_y(\tauAllan)$ decreases as $\tauAllan^{-1/2}$ until about \SI{0.2}{\second} before reaching a minimum of  \SI{47}{\pico\tesla} at about \SI{6}{\second}. $\sigma_y(\tauAllan)$ remains below \SI{60}{\pico\tesla} at \SI{20}{\second}, showing the device and magnetic environment are stable enough for measurement of slow processes such as MTB relaxation in low fields.

\begin{figure}[t]
    \centering
   
    \includegraphics[width=0.65\textwidth]{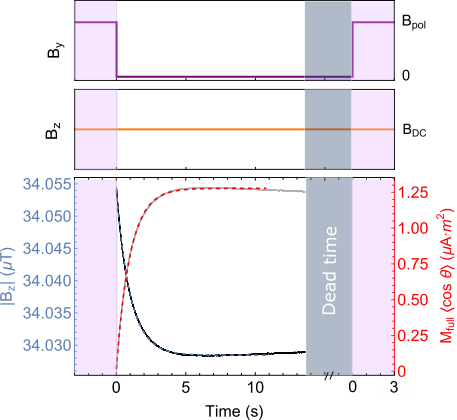}
    \caption{Top plots: Experimental sequence of applied magnetic fields. Bottom:  Bacteria alignment towards the resultant field when turning off the polarizing field, in this case $\BDC=\SI{34}{\micro\tesla}$. Black and gray curves show the field $B_z$ and the inferred magnetization (by Equation~\ref{eq:BzFromCosTheta2}) $\Mfull \langle \cos{\theta} \rangle$ respectively. The dataset consisted of 20 repeated measurements. A moving average with a window of \SI{60}{\milli\second} 
    % \mtext{mwm: what is this in time?} \rtext{60 ms since the window takes 100 elements and each element is separated 600 us} 
    was applied to each individual trace. Then, the resulting traces were averaged across the 20 measurements to obtain the final signal. Red curve fits the data with $ C + A (1-\exp[-t/\tauBact])$. Blue curve fits the data with Equations \ref{eq:BzFromCosTheta1} and \ref{eq:ExpFormB} to find $\BDC=\SI{34}{\micro\tesla}$, $ |B_\mathrm{MTB}^{(\infty)}|=\SI{25.7}{\nano\tesla}$ and $\tauBact=\SI{0.99}{\second}$. Pink colored area shows polarization field on and gray area dead time between different measurements. }
    \label{fig:BacteriaSignalvsTime}
\end{figure}

\newcommand{\mbact}{m}

\begin{figure}[t!]
    \centering
    \includegraphics[width=0.9\linewidth]{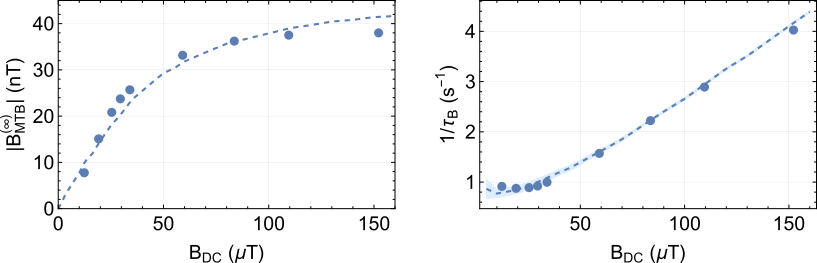}
        \caption{Measured magnetic relaxation characteristics of MTB-1. Left: bacteria signal amplitude   when changing $\BDC$. The dots represent experimental data after 20 averages and low pass filter. Right plot shows the inverse of the lifetime $\tauBact$ for different magnetic fields. In both plots dashed line represents computed relaxometry features from Eq.~\ref{eq:ThetaSDEtext} from initial condition $\theta = \pi/2$ and parameters $\mbact = \SI{0.238e-15}{\ampere\meter\squared}$, $\gamma_r = \SI{1.28e-20}{\kilo\gram\meter\squared\per\second}$, $T = \SI{32}{\celsius}$. Simulations averaged a minimum of $10^4$ realizations to compute $\langle \cos \theta \rangle$ versus time, then fit with  Eq. \ref{eq:ExpFormB} to extract best-fit equilibrium $B_\mathrm{MTB}^{(\infty)} = \lim_{t\rightarrow \infty} B_\mathrm{MTB}(t)$ and relaxation rate $\tauBact^{-1}$. Light blue colored area in the right plot indicates the error from simulation. Left plot error bars are smaller than the plot markers and are not shown.}
    \label{fig:BacteriaResult}
\end{figure}

\section{Measurement of bacterial relaxation}
\label{sec:BacteriaResults}

To study relaxation behavior of the MTB, we first apply a constant field $\mathbf{B}_\mathrm{DC} = \BDC \zhat$, i.e., along the $+z$ direction and with strength $\BDC$ in the range \SI{10}{\micro\tesla} to \SI{160}{\micro\tesla}. 
We then apply a field of $\Bpol=\SI{1}{\milli\tesla} \gg \BDC$ along the $+y$ direction using the polarizing coil for \SI{3}{\second}. This has the effect of rapidly setting the z-component of the net magnetic moment of the bacteria $M_z$ to a near-zero value.  We then switch off the strong polarizing field, leaving on $\mathbf{B}_\mathrm{DC}$, at which point $M_z$ grows toward its equilibrium value.  We take the switching off of $\Bpol$ as time zero. 

The net magnetic moment $\mathbf{M}$ of the bacteria  generates a contribution to the magnetic field at the sensor of
\begin{equation}
\label{eq:BMTB}
   \mathbf{B}\subMTB(\mathbf{r}) = \frac{\mu_0}{4\pi} \left[ \frac{3(\mathbf{M} \cdot \rhat)\rhat - \mathbf{M}}{r^3} \right],
\end{equation}
where $\mu_0$ is the permeability of free space, ${\mathbf{r} }$ is the vector pointing from the MTB centroid to the measurement point, $\rhat \equiv \mathbf{r} / r$, and $r = |\mathbf{r}|$ is the distance. Since $\BDC\gg B\subMTB$,  we can assume the magnetic field is mostly in the $z$ direction and knowing that our magnetometer is located at  ${\mathbf{r}}=-$\SI{17}{mm} $\yhat$ with \textbf{$\yhat$} the unit vector in the direction of the y-axis, the signal detected by the magnetometer is
\begin{eqnarray}
B_z(r) & = &  B_\mathrm{MTB}(r) + \BDC \label{eq:BzFromCosTheta1} \\
B_\mathrm{MTB}(r) & \equiv &  \frac{\mu_0}{4\pi} \left[ \frac{ - \Mfull \langle \cos{\theta} \rangle }{r^3} \right], \label{eq:BzFromCosTheta2}
\end{eqnarray}
where $\Mfull$ is the magnetic moment the sample would have if all the bacteria were aligned, and $\langle \cos \theta \rangle$ is the mean angle between a bacterium's polarization and the $\zhat$ direction. %The first term comes from the bacterial sample and the second is the field applied to polarize the bacteria.  
% From Figure \ref{fig:BacteriaResult}, for our conditions $B_\mathrm{MTB} \approx \SI{-45}{\nano\tesla} \times \langle \cos \theta \rangle$ at the sensor.   \mtext{mwm: I still don't get this. How do you get that from Fig. 4? }

The resulting $B_z(t)$ is observed by FID, as described in Section~\ref{Sec:Experimental system}. The polarization--relaxation procedure is repeated twenty times to obtain an averaged signal.  A typical result is shown in Figure~\ref{fig:BacteriaSignalvsTime}. The field strength $B$ decreases with time, as expected. As per common practice \cite{CHEMLA19993323, MagneticMomentExponential},
% \mtext{can we add other, more recent refs that use exponential fitting?} \rtext{\cite{MagneticMomentExponential} from 2013. Their Eq 13 is our Eq. 7. Their Fig 7 is our Fig 4}, 
we fit the measured $B_\mathrm{MTB}(t)$ with a function 
\begin{equation}
B_\mathrm{MTB}(t) = -(1-e^{-t/\tauBact}) |B_\mathrm{MTB}^{(\infty)}|
\label{eq:ExpFormB}
\end{equation}
 to obtain the equilibrium polarization $|B_\mathrm{MTB}^{(\infty)}|$ and the relaxation rate $1/\tau_{B}$, where $\tauBact$ is the alignment time.  These quantities as a function of $\BDC$ are shown in Figure  \ref{fig:BacteriaResult}. We employ this model in this section for comparison to prior results in the literature; for the largest fields used here, it is accurate to roughly  \SI{5}{\percent} of full polarization (see Section \ref{sec:Distributions}), with better accuracy at lower fields.

%\autoref{fig:BacteriaResult} (left) shows the observed $B_\mathrm{MTB}^{(\infty)}$ versus $\BDC$. 
To relate the Figure \ref{fig:BacteriaResult} (left) results to  Equation \ref{eq:BzFromCosTheta2}, which expresses $B_\mathrm{MTB}$ in terms of $\langle \cos \theta\rangle$, we first find the equilibrium value of $\langle \cos\theta \rangle$ for a range of $\BDC$ by stochastic simulation (as described in Section \ref{sec:Distributions}). We then make a single-parameter fit  of $B_\mathrm{MTB}^{(\infty)} = A \langle \cos \theta \rangle$ to the data of Fig. \ref{fig:BacteriaResult}, and find $A = \SI{-45}{\nano\tesla}$. From Eq. \ref{eq:BzFromCosTheta2}, this is the value of $-\mu_0 \Mfull /(4\pi r^3)$.

% We use the FID method described in Section~\ref{Sec:Experimental system} to observe the field magnitude $B = |\mathbf{B}_\mathrm{DC} + \mathbf{B}_\mathrm{MTB}|$, where $\mathbf{B}_\mathrm{MTB}$ is the contribution from the MTB.  

% Because $\mathbf{B}_\mathrm{DC}$ is along $\zhat$, only the $z$ component of $\mathbf{B}_\mathrm{MTB}$ contributes to the observed $B$. With the MTB located above the sensing atoms, i.e., with $\mathbf{r} = -r \yhat$, this component is, by Equation~\ref{eq:BMTB}, $\mathbf{B}_\mathrm{MTB} \cdot \zhat = -M_z \mu_0 / 4\pi r^3$. 

%The signal recorded by the magnetometer thus shows a decrease in the magnetic field during the bacterial alignment process, as $M_z$ grows due to $\mathbf{B}_\mathrm{DC}$. Although the bacterial magnetization is oriented along the $+z$ direction, the magnetic field produced at the location of the sensor is directed along $-z$, resulting in a net reduction in the measured field, as shown in Fig. \ref{fig:BacteriaSignalvsTime}.

\begin{figure}[t]
    \centering
\includegraphics[width=0.45\linewidth]{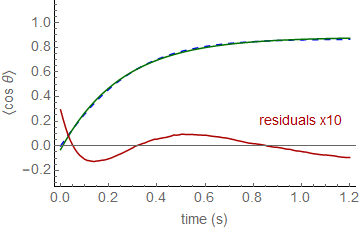}
\includegraphics[width=0.45\linewidth]{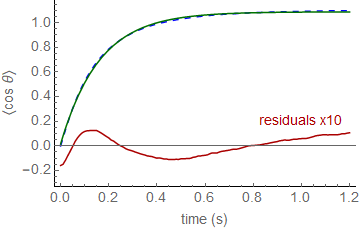} \\
\includegraphics[width=0.45\linewidth]{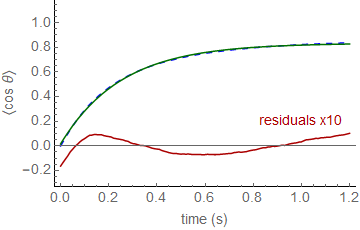}
\includegraphics[width=0.45\linewidth]{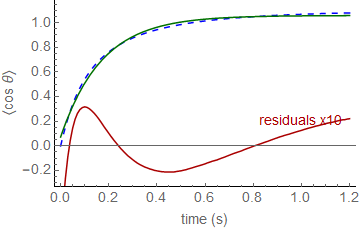}
        \caption{Non-exponential relaxation of magnetization from Equation \ref{eq:ThetaSDEtext}. Dashed blue curves show average of \SI{1e6}{} traces starting from $\theta = 0$ with $B = \SI{151.4}{\micro\tesla}$, $T = \SI{32}{\celsius}$, and  $\mbact = x_m \times  \SI{0.238e-15}{\ampere\meter\squared}$ and $\gamma_r = x_\gamma \times \SI{1.28e-20} {\kilo\gram\meter\squared\per\second}$. Green solid curve shows fit with $\langle \cos\theta \rangle = C + A (1-\exp[-t/\tauBact])$. Red curve shows residuals (simulation minus fit) times ten. Upper left graph shows a monodisperse population with $x_m = x_\gamma =1$, upper right shows a $m$-mixed population with $x_\gamma =1$,  $x_m$ log-normal distributed with $E[\ln x_m] = 0.$, $\var(\ln x_m) = 4/9$, lower left shows a $\gamma_r$-mixed population with $x_m =1$,  $x_\gamma$ log-normal distributed with $E[\ln x_\gamma] = 0.$, $\var(\ln x_\gamma) = 4/9$. Lower right shows $x_m$ and $x_\gamma$ independently log-normal distributed with $E[\ln x] = 0.$, $\var(\ln x) = 4/9$, for $x\in \{x_m,x_\gamma\}$. The results show that the relaxation in none of the cases is precisely exponential, and that dispersion of $m$ or of $\gamma$ cause similar deviations.     }
    \label{fig:NonExponentialFits}
\end{figure}

\section{Non-exponential magnetic relaxation in MTB} 
\label{sec:Distributions}

As shown in Figure~\ref{fig:BacteriaSignalvsTime}, relaxation of the mean bacterial magnetization toward equilibrium resembles an exponential approach to a steady-state value, and we have used an exponential model, Equation \ref{eq:ExpFormB}, for the fits used to generate Figure \ref{fig:BacteriaResult}. Nonetheless, angular diffusion does not in general produce exponential relaxation. Most fundamentally, this is because the angular coordinates are not Cartesian, so that the diffusion in effect occurs on a curved surface \cite{BrillingerJTP1997, YosidaAMS1949, CarlssonJPAMT2010}. A more pedestrian reason is that a single population can contain a range of values for $\mbact$ and $\gamma_r$.  Here we show that with sufficient measurement precision, the non-exponential character of MTB magnetic relaxation becomes evident, and that the departure from the exponential form can give information about the spread of $\mbact$ and $\gamma_r$ in the MTB population.

As shown in Appendix \ref{sec:BacteriaTheory}, a bacterium's polar angle $\theta$, defined relative to the field direction, obeys the stochastic differential equation 
\begin{eqnarray}
\label{eq:ThetaSDEtext}
d \theta &=& \left(\frac{B \mbact}{\gamma_r} \sin \theta +  \frac{k_B T}{\gamma_r} \cot \theta\right) dt + \sqrt{\frac{2 k_B T}{\gamma_r}} dW,
\end{eqnarray}
where $dW$ is the Wiener increment. 
%This equation does not in general lead to exponential relaxation of %$M_z \propto 
%$\langle \cos\theta \rangle$. 
We first illustrate by numerical simulation that Eq.~\ref{eq:ThetaSDEtext} produces non-exponential relaxation of $\langle \cos\theta \rangle$. We consider a bacterial population with monodisperse $\mbact$ and $\gamma_r$, and initial $\theta = \pi/2$ such that $\langle \cos\theta \rangle = 0$. This corresponds to the experimental situation immediately after the polarizing field $\Bpol$ is turned off. We average \SI{1e6}{} trajectories of Equation~\ref{eq:ThetaSDEtext}, integrated by the Milstein method in Mathematica. As shown in Figure \ref{fig:NonExponentialFits} (upper left), the resulting $\langle \cos \theta \rangle$ shows a few-percent departure from exponential behavior. The other panels of Figure \ref{fig:NonExponentialFits} show similar traces obtained by simulating the same scenario for populations with dispersion in $\mbact$, $\gamma_r$, or both. In particular, we write the mean polarization for the population as
\begin{eqnarray}
\label{eq:PopulationAverage}
\langle \cos\theta \rangle &=& \int d\mbact\, d\gamma_r f(\mbact,\gamma_r)\langle \cos\theta \rangle_{\mbact,\gamma_r},
\end{eqnarray}
where $f(\mbact,\gamma_r)$ is the probability density function of $\mbact,\gamma_r$ and $\langle \cos\theta \rangle_{\mbact,\gamma_r}$ is the mean of $\cos \theta$ for given $\mbact,\gamma_r$. 

The observed dispersion-induced departure from exponential behavior is, for most times, of opposite sign to the departure seen with a monodisperse population. In this scenario, the most nearly exponential relaxation in  fact requires some $\mbact$ and/or $\gamma_r$ dispersion to balance the effect of the curvilinear coordinates.

% A final observation is that at our level of precision we cannot reliably distinguish dispersion in $\mbact$ from dispersion in $\gamma_r$. In what follows we maintain $\mbact = m_0$

\begin{figure}[t]
    \centering    \includegraphics[width=0.48\linewidth]{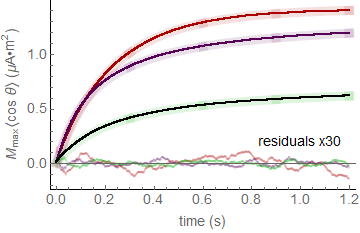}
 \includegraphics[width=0.48\columnwidth]{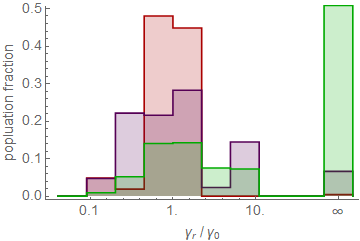}
   \caption{(Left) Relaxometry of high-concentration MTB-1 suspensions during settling and evaporation. In the presence of a magnetic field $\BDC=\SI{151.4}{\micro\tesla}$, magnetization $|M| \equiv \Mfull \langle \cos \theta \rangle$ versus time was measured as in Fig.~\ref{fig:BacteriaSignalvsTime}. Thin solid curves show data, and thick colored curves show fits, immediately after sample preparation (upper, red),  24 hours post-preparation (middle, violet) and  48 hours post-preparation (lower, green).  Curves near zero show, in corresponding colors, fit residual multiplied by thirty. Each data curve is the result of 20 averages smoothed by moving average as described in Figure~\ref{fig:BacteriaSignalvsTime} caption.  Fits are obtained as described in the text. (Right) Distribution of $\gamma_r$ found by assuming $m = m_0 \equiv \SI{0.238e-15}{\ampere\meter\squared}$ and fitting the data of Figure~\ref{Fig:Evaporation}. $\gamma_0 \equiv \SI{1.28e-20}{\kilo\gram\meter\squared\per\second}$ is the nominal damping rate.  }  
    \label{Fig:Evaporation}
\end{figure}

We can also use the shape of the relaxation to extract information about the distribution of $\mbact$ and $\gamma_r$.
From Equation \ref{eq:ThetaSDEtext}, we see that the first drift term (the magnetic drift) contains $\mbact/\gamma_r$, while the second drift term (the thermal drift) and the diffusion term contain only $\gamma_r$. Each pair $\{\mbact,\gamma_r\}$ thus leads to a distinct trajectory for $\langle \cos \theta \rangle$. In principle we can invert Equation \ref{eq:PopulationAverage} to find  $f(\mbact,\gamma_r)$. In practice, we find that increasing $\mbact$ and decreasing $\gamma_r$ have very similar effects. 

We can nonetheless extract interesting information from the relaxometry data. To illustrate this, we prepare a MTB sample and observe its relaxation at \SI{0}{\hour},  \SI{24}{\hour} and \SI{48}{\hour} as the bacterial population is concentrated by evaporation and settling, with the observed magnetization versus time shown in Figure~\ref{Fig:Evaporation}, left. We observe a progressive deviation from exponential relaxation as the evaporation proceeds; the \SI{0}{\hour} and \SI{24}{\hour} curves overlap very closely at short times but diverge afterward. The magnitude of the departure from exponential behaviour is of order $\SI{0.02}{\micro\ampere\meter\squared}$  in $M_x$ and thus of order $\SI{1}{\nano\tesla}$ in $B_\mathrm{MTB}$, well exceeding the Allan deviation values shown in Fig.~\ref{fig:MagneticSensAndStab} for integration times between \SI{0.1}{\second} and \SI{10}{\second}. This indicates that the magnetometer has sufficient sensitivity and stability to usefully measure the departure from exponential relaxation. 

To extract the parameter distribution, we perform a least-squares fit with a linear combination of calculated relaxation curves with $\mbact = 1.0$ and $\gamma_r = \exp[p] \gamma_0$, $p \in \{-2,-1.2,-0.4,0.4,1.2,2\}$ plus an offset. For the \SI{24}{\hour} and \SI{48}{\hour} fits we include also $\mbact = 1$, $\gamma_r = \infty$, to represent immobilized bacteria. In doing so, we are 1) assuming that any initial dispersion in $\mbact$ can be traded for dispersion in $\gamma_r$, since we see very similar behaviors in the simulation for $\{\mbact,\gamma_r\}$ and for $\{x \mbact,x \gamma_r\}$ for $x\sim 1$, and 2) assuming that over the course of the experiment $\gamma_r$ may change, while $\mbact$ is stable. 

The resulting weights are shown in Figure~\ref{Fig:Evaporation} (right graph), and the fitted curves and residuals are shown in Figure~\ref{Fig:Evaporation} (left graph). The results show an initially narrow distribution of $\gamma_r$, evolving into a broader distribution, with a growing immobile component $\gamma_r = \infty$.  While the microscopic origin of this behavior is not yet clear, it is plausible that the local bacterial environments and thus drag coefficients become more varied as a result of evaporation and settling. It is also directly observed that a non-negligible fraction of the MTB sample was stuck to the walls of the container after \SI{48}{\hour}. 
%In addition, this effect also lead to slower alignment dynamics.

% \begin{figure}[t!]
%     \centering
% \includegraphics[width=0.6\columnwidth]{images/GammaDistPlot.png}
%         \caption{Distribution of $\gamma_r$ found by fitting the data of Figure~\ref{Fig:Evaporation}. $\gamma_0 = \SI{1.28e-20}{\kilo\gram\meter\squared\per\second}$ is the nominal damping rate.  
%        \rtext{maybe we can do Fig 5 and 6 combined?} }
%     \label{fig:Distributions}
% \end{figure}

\section{Conclusions}
\label{sec:Conclusions and outlook}

We have demonstrated optically-pumped magnetometery with a $\SI{}{\milli\meter^3}$-scale atomic vapor volume, using a Pound-Drever-Hall optical-cavity readout scheme  described in \cite{HernandezPRAppl2024}. Using an FID method, the PDH-OPM achieves, in fields up to \SI{160}{\micro\tesla}, an instability below \SI{60}{\pico\tesla} for measurement durations between \SI{1}{\second} and \SI{20}{\second} with minimal shielding (a single-layer mu-metal shield, open to allow rapid sample access).  This sub-ppm stability allows measurement of slow signals from weak sources.  We apply the PDH-OPM to the measurement of relaxation in a population of the magnetotactic bacterium \textit{Magnetospirillum gryphiswaldense}, strain MSR-1, and quantify the evolution of the distribution of damping coefficients as a consequence of concentration during settling and evaporation. We leave for future work detailed studies of the mechanism behind this observation. To our knowledge, this is the first time such magnetic inhomogeneities and long-term relaxation deviations have been directly observed. These results highlight the capability of compact atomic magnetometers to image and quantify collective magnetic dynamics in biological magnetic systems, particularly in environments where optical techniques are hindered by scattering or opacity.

While the PDH-OPM operates at the photon–shot–noise limit, aside from technical noise peaks primarily at harmonics of the mains frequency, there remains scope for improving the measurement sensitivity by increasing the optical cavity finesse, raising the PDH modulation frequency, and achieving critical coupling during cavity readout. The current undercoupled cavity limits the efficiency of the optical readout. Beyond improving the PDH-OPM itself, implementing a gradiometric configuration in future designs would significantly enhance magnetic background noise rejection.

\begin{appendices}

\section{Simulating bacterial angular diffusion}
\label{sec:BacteriaTheory}

\newcommand{\bxi}{\boldsymbol{\xi}}
Considering a single bacterium, the angular motion is described by the Langevin equation \cite{CoduttiPLOS2019}
\begin{eqnarray}
\label{eq:CoduttiLE}
\frac{d}{dt} \mathbf{e} &=& \frac{1}{\gamma_r } [\mathbf{T}_\mathrm{ext} + \sqrt{2 k_B T \gamma_r} \bxi_r]\times \mathbf{e},
\end{eqnarray}
where $\mathbf{e}$ is a unit vector indicating the bacterial orientation,  $\gamma_r$ is the rotational friction coefficient, which for a cylindrical bacterium of diameter $d$ and length $L$ in a medium of viscosity $\eta$ is $ \gamma_r\approx\frac{\pi \eta L^3}{3}\left[\ln{\left(\frac{L}{d}\right)-0.662-0.92\frac{d}{L}}\right]^{-1}$,  $\mathbf{T}_\mathrm{ext} = \mbact \mathbf{e} \times \mathbf{B} $ is the torque due to the magnetic field, $k_B T$ is the mean thermal energy and $\bxi_r$ is a 3-vector of independent unit white noises. This last term describes an isotropic, delta-correlated torque of thermal origin. 

%The external torque is given by 
% \begin{eqnarray}
% \mathbf{T}_\mathrm{ext} &=& \mbact \mathbf{e} \times \mathbf{B}. 
% \end{eqnarray}

The above SDE clearly preserves the length of $\mathbf{e}$. It describes biased diffusion on a unit sphere, which has been studied in a variety of contexts \cite{BrillingerJTP1997, YosidaAMS1949, CarlssonJPAMT2010}. We can recast Eq.~\ref{eq:CoduttiLE} in the form 
\begin{eqnarray}
\label{eq:SDEsphere}
\frac{d}{dt} \mathbf{e} &=& \frac{1}{\gamma_r} \mathbf{T}_\mathrm{ext} \times \mathbf{e} + \sigma (\mathbbm{1} - \mathbf{e} \mathbf{e}^T ) \cdot \bxi_r,
\end{eqnarray}
where $\mathbbm{1}$ is the identity  and $\sigma = \sqrt{2 k_B T/ \gamma_r}$ is the diffusion coefficient. In polar coordinates $\theta, \phi$ with $\theta = 0$ when $\mathbf{e}$ is along the $\mathbf{B}$ direction, the first term above contributes a $\theta$ drift rate $B \mbact \gamma_r^{-1} \sin\theta$, while second term contributes an isotropic diffusion. 

This problem is reviewed in detail in Brillinger \cite{BrillingerJTP1997}. A useful result for us  \cite[Eq. 3.5]{BrillingerJTP1997} is that for $\mathbf{T}_\mathrm{ext} =0$, the polar coordinates obey the Itô stochastic differential equations %\cite[Eq. 3.5]{BrillingerJTP1997} 
\begin{eqnarray}
\label{eq:BrillingerThetaSDE}
d \theta &=& \frac{1}{2} \sigma^2 \cot \theta dt + \sigma dW_{\theta} \\
\label{eq:BrillingerPhiSDE}
d \phi &=&  \sigma \csc \theta  dW_{\phi},
\end{eqnarray}
where $dW_{\theta}$ and $dW_{\phi}$ are independent Wiener increments. The $\theta$ drift contribution $\cot\theta$ is a result of the curvature of the spherical surface, and reflects the increased phase space density near the equator, where $\theta = \pi/2$. 

Reintroducing the drift term, we find $\mathbf{T}_\mathrm{ext}$ contributes a $\theta$ drift rate $B \mbact \gamma_r^{-1} \sin\theta$, at which point we have  the Itô stochastic differential equation for magnetically-biased angular diffusion
\begin{eqnarray}
\label{eq:ThetaSDE2}
d \theta &=& \left(\frac{B \mbact}{\gamma_r} \sin \theta +  \frac{k_B T}{\gamma_r} \cot \theta\right) dt + \sqrt{\frac{2 k_B T}{\gamma_r}} dW.\hspace{9mm}
\end{eqnarray}
Numerical calculation shows that in the long-time limit this reproduces the Langevin formula for the mean polarization
\cite{RevModPhys.96.021001}
\begin{equation}
\langle \cos\theta \rangle =\coth \frac{m B}{k_BT}- \frac{k_BT}{m B}.
\label{Eq:cosx}
\end{equation}
We note that this formula derives from equilibrium thermodynamics and the geometry of the angular coordinates, and as such is an independent check on the dynamical model in Eq. \ref{eq:ThetaSDE2}.

% \section{Non-exponential relaxation of MTB}
% \label{sec:Distributions}

\end{appendices}

\bmhead{Acknowledgements}

We thank Danny Yosmar Villanueva-\'{A}lvaro, Mar\'{i}a Luisa Fern\'{a}ndez Gubieda Ruiz, Ana Abad D\'{i}az de Cerio and Ana Garc\'{i}a Prieto for providing magnetotactic bacteria and helpful discussions. 
This work has been supported by European Commission projects Field-SEER (ERC 101097313), OPMMEG (101099379) and QUANTIFY (101135931); Spanish Ministry of Science MCIN projects SAPONARIA (PID2021-123813NB-I00),  SALVIA (PID2024-158479NB-I00), and MARICHAS (PID2021-126059OA-I00), ``NextGenerationEU/PRTR.'' (Grant FJC2021-047840-I) and ``Severo Ochoa'' Center of Excellence CEX2019-000910-S;  Generalitat de Catalunya through the CERCA program,  DURSI grant No. 2021 SGR 01453 and QSENSE (GOV/51/2022).  Fundaci\'{o} Privada Cellex; Fundaci\'{o} Mir-Puig. MHR acknowledges support from Ayuda PRE2021-098880 financiada por MCIN/AEI/ 10.13039/501100011033 y por el FSE+. VGL acknowledges financial support from European Union NextGenerationEU PNRR MUR project MAPPIQS, NQSTI Spoke 9 - CUP E63C22002180006 and from the Italian Ministry of
University and Research (MUR) projects “Rita Levi Moltancini”—Bando 2021, AQUSENS and "Budget MIUR
- Dipartimenti di Eccellenza 2023 - 2027” (Law 232,
11 December 2016) - Quantum Sensing and Modelling
for One-Health (QuaSiModO).


\begin{appendices}




\end{appendices}


\bibliography{sn-bibliography}% common bib file

\end{document}